\documentstyle[12pt]{article}
\begin{document}
\renewcommand{\thefootnote}{\arabic{footnote}}
\setcounter{footnote}{0}
\begin{titlepage}
\renewcommand{\thefootnote}{\fnsymbol{footnote}}
\makebox[2cm]{}\\[-1in]
\begin{flushright}
\begin{tabular}{l}
hep-ph/9505317\\
May 1995
\end{tabular}
\end{flushright}
{}
\vspace*{2.0cm}
\begin{center}
{\large\bf
QCD RENORMALONS AND HIGHER TWIST EFFECTS
}

\vspace{1.3cm}

V.M.\ Braun\footnote{On leave of absence from St.\ Petersburg
Nuclear Physics Institute, 188350 Gatchina, Russia.}


{\em DESY, Notkestra\ss e 85, D--22603 Hamburg, Germany}

\vspace{6.5cm}

{\bf Abstract\\[5pt]}
\parbox[t]{\textwidth}{\small
I give a short review of the relation  of infrared renormalons in QCD
and higher twist effects, with the emphasis on possible applications.
In particular, I present estimates of renormalon-induced
uncertainties in deep inelastic sum rules and explain how the
renormalons can potentially  be used to unravel the structure of
nonperturbative effects in complicated situations
and to indicate possible systematic sources of large perturbative
corrections.
}

\vfill

\end{center}

\centerline{\it to appear in the Proceedings of the XXXth Rencontres
de Moriond}
\centerline{\it ``QCD and High Energy Hadronic Interactions''}
\centerline{\it Les Arcs, France, March 1995}

\end{titlepage}
\renewcommand{\thefootnote}{\arabic{footnote}}
\setcounter{footnote}{0}
\newpage
\bigskip

{\large\bf 1.}\qquad
The modern precise data on ``hard''
processes in QCD require a theoretical description to power-like accuracy.
A clear example is provided by the Gross-Llewellyn Smith sum rule in the
deep inelastic scattering, where the higher-twist (HT) $1/Q^2$ correction
produces a major uncertainity in determination of $\alpha_s$
\cite{GLS}. For a generic physical observable dominated by short distances
one expects a theoretical prediction of the form, schematically,
\begin{equation}\label{general}
R(Q) = R_{tree}\Big[1+r_1\frac{\alpha_s}{\pi}
      +r_2\left(\frac{\alpha_s}{\pi}\right)^2+\ldots\Big]
      +\sum_s\frac{H_s}{(Q^2)^s}\,,
\end{equation}
where $\alpha_s=\alpha_s(Q)$ and $H_s$'s are dimensionful nonperturbative
parameters (with dimension $2s>0$) describing the HT corrections.
A conceptual problem, which I am going to review in this
talk, is that the discrimination between perturbative corrections and HT
contributions is ambiguous.
This will imply that it is not possible to attribute a fully quantitative
meaning to power-suppressed corrections, unless some prescription is used
to sum the perturbative series. On the other hand, ambiguities in summation
of the perturbative series can serve to indicate  which powers
of $1/Q$ are required in the sum in (\ref{general}).

Assuming for simplicity a (euclidian) quantity dominated by a large scale $Q$,
the leading-order correction involves the gluon exchange with
the large virtuality of order $Q$.
 Progressing to a higher order $n$
the average gluon virtuality still remains  proportional to $Q$,
$k\sim a_n Q$ where $a_n$ is a certain coefficient, simply because
there are no more dimensionful parameters. However,
 the $a_n$ can (and do) decrease with $n$, so that in
very high orders $n$ such that $a_n \sim Q/\Lambda_{\rm QCD}$ the perturbative
calculation fails.
An inspection shows
that the most dangerous Feynman diagrams are those related to running of
the QCD coupling, in which the average virtuality decreases exponentially
$a_n\sim \exp[-n/s]$, where $s$ is some number.
It is possible to show that this intervention of infrared (IR) regions reveals
itself in a rapid -- factorial -- increase of perturbative coefficients
in high orders \cite{renorm}:
\begin{equation}\label{highorders}
   R(Q) = R_{tree}\sum_n r_n \alpha_s^n(Q);\qquad r_n\sim (\beta_0/s)^n n!
\end{equation}
where $\beta_0=(11-2/3 n_f)/(4\pi)$.
The physical origin of the large coefficients is simple:
the  gluon exchange with virtuality $k$ involves the QCD coupling
at this scale $\alpha_s(k)$. However, the perturbative expansion
(\ref{highorders}) is assumed to be in powers of $\alpha_s(Q)$ at the
scale of the external momenta. Thus,  we get large coefficients
simply by reexpressing $\alpha_s(k\sim \exp(-n/s)Q)$ in terms of $\alpha_s(Q)$.

A factorial growth of perturbative coefficients means that the perturbative
series is at best an asymptotic series: the fixed order contributions
$r_n\alpha_s^n$ decrease at small $n$, reach a certain minimum value at
 $n=n_0\sim 1/\alpha_s$,  but then again start to increase and
blow up. This means that a perturbative calculation only makes sense
up to the order $n_0$, and the accuracy of this calculation, or, equivalently,
the effect of the ``tail'' with $n>n_0$ is of the order of the minimum
term
\begin{equation}\label{tail}
     \sum_{n=n_0}^\infty r_n\alpha_s^n(Q) \sim r_{n_0}\alpha_s^{n_0}(Q)
    \sim \exp[-s/(\beta_0\alpha_s(Q))]
    \sim \left(\frac{\Lambda^2_{\rm QCD}}{Q^2}\right)^s\,,
\end{equation}
where I used the asymptotic form of the coefficients (\ref{highorders}) and
the one-loop formula $\alpha_s(Q)=1/(\beta_0\ln(Q^2/\Lambda^2))$.
In the common jargon, the divergences of perturbative expansions are called
``renormalons'' \cite{renorm}, and I will refer to the
ambiguity in the summation
of the series (\ref{tail}) as to the ``renormalon ambiguity''.

The deficiency of the perturbation theory has a profound reason, indicating
that calculation of physical quantities to power-like accuracy requires
taking into account nonperturbative effects. The renormalon ambiguities
 must be compensated by ambiguities in the
HT corrections complementing truncated perturbative expansions.
Hence, the required values of $s$ in (\ref{highorders}), (\ref{tail})
have to be in one-to-one correspondence to the required HT
corrections in (\ref{general}). This implies that the required powers
of $1/Q$ can  {\em in principle}
be determined from purely perturbative calculations.
In practice,
already a simple approximation (referring to the $1/N_f$ expansion)
usually recognizes  all power-suppressed corrections which are required
by the Operator Product Expansion (OPE).

\bigskip

{\large\bf 2.}\qquad
The relation of IR renormalons and the OPE has been studied in much detail
\cite{Mueller1,David,ZAK92,BEN93a} for the polarization operator of
vector currents.
In this case the perturbative series is complemented by the contribution
of the gluon condensate \cite{SVZ}
\begin{equation}\label{GG}
  Q^2\frac{d}{dQ^2}\Pi(Q^2) = 1 +\frac{\alpha_s(Q)}{\pi} +\ldots
   -\frac{1}{6Q^2}\langle g^2 G^2 \rangle + O(1/Q^6)
\end{equation}
By an explicit calculation \cite{BEN93a} it has been shown that the
perturbative series diverges, producing an ambiguity $O(1/Q^4)$
(s=2,3,\ldots in
(\ref{tail})). On the other hand, numerical value of the gluon
condensate cannot be determined to better accuracy than of order
$\Lambda_{\rm QCD}^4$ because of the quartic power divergence.
These uncertainties must mutually cancel, since they only arise because of
our (illegal) attempt to separate perturbative and nonperturbative effects,
and an immediate question is whether one can organize the expansion in
such a way that this problem does not appear.
A possible solution is suggested by the Wilson OPE,
which in fact is not designed to separate perturbative and nonperturbative
effects, but rather to separate contributions of small and large distances.
Following this logic literally, we would subtract from the perturbative
answer the contribution of small virtualities $k^2 <\mu^2$ (where $\mu$ is
 of order 1 GeV), and add it to the gluon condensate as
a perturbative contribution of order $\mu^4$. The premium is that
the separation between the subtracted perturbative answer and the power
suppressed contribution of the condensate is now unambigous: since the
IR region is deleted, the
perturbative expansion is not plagued by factorially large coefficients and
the cancellation
of ambiguities between perturbative and nonperturbative contributions
to the condensate becomes implicit. The price to pay, however, is that
both perturbative
and condensate contributions depend explicitly of the scale $\mu$.
This dependence may be strong: elimination of the renormalon ambiguity of
order $\Lambda^4_{\rm QCD}$ requires reshuffling of a much bigger
contribution of order $\mu^4$.
Thus, usefullness of this rearreangement should be judged by
the practical gain: if renormalon ambiguities
 are numerically much
smaller than the phenomenological estimates for the condensates,
it is hardly   reasonable to  eliminate the
ambiguity at the cost of a large $\mu$ dependence.
In fact, it is easy to see that the whole idea to add HT
corrections to the perturbative expansions, in which we only know a few
first terms and do not see any sign of the factorial divergence
(and thus expect that we have not yet reached the minimum term), implicitly
 {\em implies} that the ``true nonperturbative'' HT  corrections
are much bigger than
the minimum term in the perturbative series, and thus much bigger than
the uncertainity in its summation.

In this respect it is important that semiquantitative estimates of
ambiguities in the summation of the perturbative series can be
worked out. For the particular case of the polarization operator using the
results of \cite{BEN93a} I get
\begin{equation}\label{GGG}
  Q^2\frac{d}{dQ^2}\Pi(Q^2) = \Bigg[1 +\frac{\alpha_s(Q)}{\pi} +\ldots
   \pm \frac{0.002 - 0.02\,\mbox{\rm GeV}^4}{Q^4}\Bigg]
   - \frac{(0.08 -0.12)\,\mbox{\rm GeV}^4}{Q^4}
   + O(1/Q^6)
\end{equation}
where the first number (with an error bar) is an estimate of the renormalon
uncertainty\footnote{The simplest way
to make this estimate is by the minimum term in the perturbative expansion.
I use a somewhat more analytic method and give an imaginary part
(divided by $\pi$) of the Borel transform.},
and the second number refers to the phenomenological value of the gluon
condensate \cite{SVZ}. The large range of values for the renormalon ambiguity
is due to the fact that it is proportional to the fourth power of
$\Lambda_{QCD}$ and the uncertainty in the latter is amplified.
It is seen that the ambiguity is in fact
 insignificant compared to a 50\%
 error in the phenomenological value of $\langle g^2G^2\rangle$.
This fact was recognized long ago (see, e.g. \cite{SHI93}) and is
one of the starting points of the QCD sum rules, where
the QCD renormalons are ignored as being (supposedly)
numerically insignificant.

For the  GLS and Bjorken sum rules,
assuming the range of values  $\alpha_s(Q^2=3$ GeV$^2)=0.24 - 0.29$, as
suggested by the new CCFR data \cite{GLS},
I get
\begin{eqnarray}
\mbox{\rm GLS} &=& 3\Bigg\{\Bigg[1 -\frac{\alpha_s(Q)}{\pi} +\ldots
   \pm \frac{0.02 - 0.07\,\mbox{\rm GeV}^2}{Q^2}\Bigg]
   - \frac{(0.1\pm 0.03)\,\mbox{\rm GeV}^2}{Q^2}
   + O(1/Q^4)
\\
\mbox{\rm Bj} &=& \frac{1}{6}\Bigg\{g_A\Bigg[1 -\frac{\alpha_s(Q)}{\pi} +\ldots
   \pm \frac{0.02 - 0.07\,\mbox{\rm GeV}^2}{Q^2}\Bigg]
   - \frac{(0.09\pm 0.06)\,\mbox{\rm GeV}^2}{Q^2}
   + O(1/Q^4)
\end{eqnarray}
where the numerical values of the HT $1/Q^2$ corrections are taken from the
QCD sum rule calculations \cite{BBK}. In this case the renormalon ambiguities
are roughly factor two smaller than the ``true'' HT effects.
Note that the renormalon ambiguities are determined by
divergences in the coefficient functions in the OPE and
are universal in the sense that they do not depend
on the particular target, while ``geniune'' HT corrections reflect
correlations between partons in the target (nucleon) and
are process-dependent.

Finally, it has been shown \cite{BIG94,BB94} that the pole mass of a
heavy quark is affected by IR renormalons already at the level of
$1/m$ corrections, which do not allow to determine
it from the relation to the mass defined at short distances to the accuracy
better than
\begin{equation}
  m^{\rm pole} = \overline{m}(\overline{m})\Bigg[
1 + \frac{4}{3}\frac{\alpha_s(m)}{\pi} +\ldots\Bigg] \pm 100\,
    \,\mbox{MeV}
\end{equation}
This number should be compared to the ``true'' nonperturbative mass difference
between the heavy meson and the heavy quark, which is estimated to be of order
$\bar\Lambda \equiv m_B-m_b \sim 300-500$ MeV \cite{lambdabar}.

To summarize, the ``renormalon problem'' of the separation
of perturbative and nonperturbative contribution can be more or less important
in each practical case, depending on which method is used to estimate the
HT contribution, and what accuracy is claimed.
Existing calculations of the HT corrections
do not pretend to an accuracy better than of order 30-50\%
which is larger (or of order) of the estimated renormalon uncertainty,
so that the latter can be ignored.
However,
to improve these  predictions  one
has to treat the renormalon problem explicitly.

\bigskip

{\large\bf 3.}\qquad
The IR renormalons have received much attention recently  because of
their potential to expose
power-like corrections:
the $\sim 1/Q^{2s}$ uncertainty in summation of the perturbative series
should be interpreted as indicating presence of the $\sim 1/Q^{2s}$
nonperturbative correction in the particular physical observable.
It should be stressed that this argument cannot compete with the
OPE if the latter is applicable: attempts to ``check'' or to ``disprove''
OPE using the renormalon arguments are essentially meaningless.
It is precisely the agreement with the OPE
in its traditional domain which serves as main justification
of the hope that the IR renormalons may provide a nontrivial
information about nonperturbative effects in a more general situation,
since this  technique relies on a purely
perturbative analysis and can be relatively easily implemented.

Despite the complicated terminology, the idea of this
application is simple, and is that IR renormalons in fact probe IR sensitivity
of perturbative diagrams to power-like accuracy.
This  can be made explicit
using the result of Refs.\cite{BBZ,BBB95}:
for IR-safe inclusive quantities there is a one-to-one correspondence
between IR renormalons and {\em nonanalytic} terms in the expansion
of corresponding Feynman diagrams in powers of some  IR regulator
like a small gluon mass $\lambda$. In particular, existence of the
$\sqrt{\lambda^2}$ term
in this expansion signals existence of the $1/Q$ uncertainty
in summation of the perturbative series, and this necessitates the $1/Q$
nonperturbative correction. The  $\lambda^2\ln\lambda^2$ term implies
the $1/Q^2$ correction, etc. For illustration, consider the following
result \cite{BBB95} for the polarization operator (\ref{GG})
calculated with a small gluon mass:
\begin{eqnarray}\label{Dexpand}
 Q^2\frac{d}{dQ^2}\Pi(Q^2)
    &=&1+\frac{\alpha_s}{\pi}\Bigg\{
1-\left[\frac{32}{3}-8\zeta(3)\right]\frac{\lambda^2}{Q^2}-
\left[2 \ln (Q^2/\lambda^2)+\frac{20}{3}-8\zeta(3)\right]
\frac{\lambda^4}{Q^4}
\nonumber\\&&{}+O\left(\lambda^6\ln^2 \lambda^2/Q^2\right)\Bigg\}\,.
\end{eqnarray}
Note that there are no terms  $\sim \lambda^2\ln\lambda^2$
(analytic terms like $\sim \lambda^2$ are not related to the IR region)
and the first nonanalytic correction is of order
$\sim \lambda^4\ln\lambda^2$, indicating a potential nonperturbative
 contribution of order $1/Q^4$, in agreement with the OPE (\ref{GG}).
Note that absence of terms $\sim \ln \lambda^2$ is the result of
celebrated Bloch-Nordsieck cancellations between contributions of real
and virtual emission, and absence of certain
power-like corrections (alias renormalon ambiguities) can be
formulated as extension of Bloch-Nordsieck cancellations to power-like
accuracy \cite{BBZ}.

The search of nonperturbative effects using IR renormalons  has been
been most fruitful in  heavy quark decays \cite{BIG94,BB94,BBZ}.
One finds for the b-quark pole mass
\begin{equation}\label{pole2}
 m_b^{\rm pole} = \overline{m}_b(\overline{m}_b)\Bigg[
1 + \frac{4}{3}\frac{\alpha_s(m)}{\pi} -
   \frac{2}{3}\alpha_s \frac{\sqrt{\lambda^2}}{m_b}+\ldots\Bigg]
\end{equation}
The term $\sim \sqrt{\lambda^2}$ indicates presence of an $1/m$
nonperturbative correction (which is not related to matrix element
of any local operator and cannot be found using the usual OPE).
On the other hand, the B-meson total semileptonic inclusive width equals
to the same accuracy  (for simplicity I give the answer for
$b\rightarrow u e\nu$ transitions, that is for massless quark in the final
state)
\begin{equation}
 \Gamma(B\to X_u e\nu) = \frac{G_F^2}{192 \pi^3} (m_b^{\rm pole})^5
\Bigg[1-2.41 \frac{\alpha_s(m)}{\pi}
  + \frac{10}{3}\alpha_s \frac{\sqrt{\lambda^2}}{m_b}+\ldots\Bigg]
\end{equation}
The $\sim \sqrt{\lambda^2}$ correction is again  present,
but is cancelled exactly if the pole mass of the b-quark is eliminated
in terms of the $\overline{\rm MS}$ running mass using (\ref{pole2}).
This cancellation presents the result of a prime physical importance:
inclusive decay widths of heavy particles do not contain nonperturbative
corrections of order $1/m$ if they are expressed in terms of mass parameters
defined at short distances \cite{BIG94,BBZ}.

Further applications of IR renormalons to the study of nonperturbative
efects in resummed  cross sections will be reviewed by G. Korchemsky
\cite{KOR}.

\bigskip

{\large\bf 4.}\qquad
One more idea which has emerged from studies of the QCD renormalons is
 that the  Feynman diagrams related to
running of the strong coupling, whose low-momentum regions produce
renormalons, may give  dominant contributions to perturbative coefficients
in intermediate orders and can be identified and resummed
\footnote{For this
purpose one also needs to consider the so-called
ultraviolet renormalons, which are related to factorial divergences of
perturbative series arising from contributions of momenta $k \gg Q$ in
Feynman diagrams
and which I did not discuss in this talk.}.
This can be considered as a natural generalization
of the proposal by Brodsky-Lepage-Mackenzie (BLM) \cite{BRO83} to eliminate
all dependence on the QCD $\beta$-function from coefficients
of the perturbation theory by adjusting the scales of the coupling separately
in each order. Several examples show that
the second order perturbative coefficient is significantly reduced by
this rearrangement. In high orders one again expects
a considerable reduction of coefficients since IR renormalons are made implicit
(they are hidden in uncertainties of the BLM scales \cite{BB94b}). Thus,
it is natural to speculate that the BLM-improved perturbation theory has
smaller coefficients to all orders.

The corresponding resummation of running coupling effects to all orders
has been proposed in \cite{BB94b,Nnew,BBB95}.
 Formally, this approach allows to
calculate all perturbative corrections of order $\beta_0^n\alpha_s^{n+1}$
and $\beta_1\beta_0^{n-2}\alpha_s^{n+1}$ which can be traced by contributions
of fermion bubble insertions into the single gluon line.
The resummation of running coupling effects is relatively simple and
 is probably phenomenologically relevant, although
the dominance of these correctons is a conjecture, which can only be justified
{\em a posteriori} by comparing to exact calculations.
At present the corresponding calculations have been done for the
heavy quark pole mass \cite{BB94b,BBB95}, the Adler function and
 the $\tau$-lepton hadronic width \cite{Nnew,BBB95}, and for
the exclusive \cite{Nnew} and inclusive \cite{BBB295} B-decays.
Our results for the $\tau$ decay
show that the value of $\alpha_s(m_\tau)$ extracted from these data is
probably overestimated, and also give some indication that the
commonly used
resummation of $\pi^2$ corrections is
disfavoured in high orders, see \cite{BBB95} for details.
The resummation of $\beta_0^n\alpha_s^{n+1}$ corrections in B-decays
 and its relevance
for the extraction of $V_{cb}$  will be addressed
in her talk by P. Ball \cite{BALL}.

\bigskip

{\large\bf 5.}\qquad
To summarize, I repeat that the QCD perturbation theory
is divergent and does not allow to give quantitative predictions to
 power-like accuracy, unless it is complemented by explicit
nonperturbative (HT) corrections.
In turn, the HT corrections are by themselves
ill-defined. The corresponding ambiguities have to be in one-to-one
correspondence to the ambiguities of perturbation theory and must cancel
in the sum. For practical cases of the GLS and Bj sum rules the
corresponding ambiguities are probably  a factor 2 below
the ``true'' HT corrections.
The increasing interest in IR renormalons is trigged by hopes that
they can help to investigate the structure of
nonperturbative corrections in rather general situations, and
to find physical observables with extended IR stability
(to power accuracy), which are less sensitive to nonperturbative effects.
Another hope is to get estimates for higher-orders of perturbation
theory, combining the information about the calculated low orders and
about the renormalons in very high orders. Both directions are
interesting and worth efforts.

As a final remark, let me say that nonperturbative effects in QCD are not
reduced to renormalons. In particular, not all nonperturbative effects
can be traced by divergences of the perturbation theory, and also
by no means the renormalons can be used to define QCD {\em nonperturbatively}
in the region where the coupling becomes strong.
Equally, my talk cannot pretend to cover all aspects of
QCD renormalons --- for example, I ignored ultraviolet renormalons which
deserve a special discussion. I refer the readers to the review \cite{MUE92}
and original papers for the discussion of issues which I was not able to
touch here. I thank  the organizers of this conference
for the invitation, and gratefully acknowledge a very rewarding collaboration
with P. Ball, M. Beneke and V. Zakharov on
the subjects related to this talk.

\end{document}